\documentclass[reprint,groupedaddress,floatfix]{revtex4-2}

\usepackage[pdftex]{graphicx}
\usepackage{color}
\graphicspath{{./},{./figs/},{./supplement/}}

\usepackage{bm}
\usepackage{amsmath}
\usepackage{braket}
\usepackage{hyperref}

\newcommand{\Figref}[1]{Fig.~\ref{#1}}

\newcommand{\Eqref}[1]{Eq.~(\ref{#1})}
\newcommand{\Eqsref}[1]{Eqs.~(\ref{#1})}
\newcommand{\Braref}[1]{(\ref{#1})}

\begin{document}
\title{Collective Motion of Quincke Rollers with Fully Resolved Hydrodynamics}

\author{Shun Imamura$^1$}
\author{Kohei Sawaki$^1$}
\author{John J. Molina$^1$}
\author{Matthew S. Turner$^{1,2}$}
\author{Ryoichi Yamamoto$^1$}
\email{ryoichi@cheme.kyoto-u.ac.jp}
\affiliation{
$^1$Department of Chemical Engineering, Kyoto University, Kyoto 615-8510, Japan\\
$^2$Department of Physics, University of Warwick, Coventry CV4 7AL, UK}

\date{\today}

\begin{abstract}
A Quincke roller is a unique active particle that can run and tumble freely on a flat plate due to the torque generated by a uniform DC electric field applied perpendicular to the plate\cite{bricard2013emergence}.
A system involving many such particles exhibits a variety of collective dynamics, such as the disordered gas, polar liquid, and active crystal states\cite{bricard2013emergence, bricard2015emergent, lu2018pair, karani2019tuning, geyer2019freezing, mauleon2020competing, zhang2021persistence, liu2021activity}.
We performed direct numerical simulations of a three-dimensional system containing many self-rotating particles to explicitly resolve the hydrodynamic interactions among rotating particles.
The collective motion depends on the magnitude of the dipole moments induced on the dielectric particles, the area fraction of particles, and the strength of interparticle attraction.
We find that the highly ordered polar liquid state is destabilized by the hydrodynamic interaction between rotating particles at high densities: the near-field lubrication interaction becomes dominant over far-field effects as the interparticle separation becomes shorter.
\end{abstract}


\maketitle

\begin{figure}[htbp]
  \centering
  \includegraphics[clip, width=8.0cm]{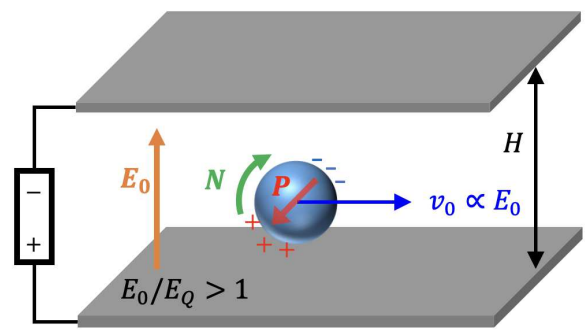}
\caption{
A schematic illustration of a Quincke roller. When a DC electric field $E_{0}$ greater than a threshold $E_{Q}$ is applied to a dielectric particle located near the lower of two plates in an electrolyte fluid, the particle spontaneously rotates and is propelled to move on the plate due to an electrohydrodynamic instability called the Quincke effect.
Here, $\bm{P}$ is the induced electric dipole moment, $\bm{N}$ is the torque exerted due to the Quincke effect, $v_{0}$ $(\propto E_{0})$ is the propulsion velocity, and $H$ is the distance between the lower and upper electrode plates.
}
  \label{fig: system}
\end{figure}

When a uniform DC electric field is applied to a dielectric particle in an electrolyte fluid, spontaneous rotation of the particle can be observed due to an electrohydrodynamic instability known as the Quincke effect \cite{quincke1896ueber}  (see \Figref{fig: system}).
A Quincke roller is a self-propelled particle whose propulsion is driven by the hydrodynamic traction force exerted on a particle rotating due to the Quincke effect near a substrate \cite{bricard2013emergence}.
The propulsion velocity of the roller is proportional to the angular velocity, which can be controlled via the magnitude of the external electric field \cite{pannacci2007insulating, pradillo2019quincke}.

A variety of distinct regimes have been observed in experimental studies, including gas, polar liquid, active solid and cluster states \cite{bricard2013emergence, lu2018pair, karani2019tuning, geyer2019freezing, mauleon2020competing, zhang2021persistence, liu2021activity}, as well as an order--disorder transition between the polar liquid and gas states \cite{bricard2013emergence}, mobility enhancement due to paired-up states \cite{lu2018pair}, and solidification at high density \cite{geyer2019freezing}.
Numerical studies have also been conducted, mostly using two-dimensional particle-based models. These have had partial success in reproducing some experimental observations \cite{bricard2013emergence, mauleon2020competing}.
The far-field hydrodynamic interaction (HI) between Quincke rollers has been considered using Stokesian dynamics \cite{swan2007simulation, balboa2017hydrodynamics, driscoll2017unstable}.
A continuum model has also been used to investigate the linear stability of the liquid--solid flocking transition \cite{bricard2013emergence, geyer2019freezing}.
In previous numerical studies the near-field HI has been greatly simplified, even for systems at high densities where a more faithful treatment of the lubrication effect between rotating particles becomes essential.

In this paper, we report direct numerical simulations (DNS) of a three-dimensional system containing many self-rotating particles in which we explicitly resolve the HIs between them.
We first present the details of our numerical implementation method. Then we test the validity of the present DNS method by comparing our results with theoretical predictions for a single particle system.
The role and importance of the near-field hydrodynamics is then examined for two- and three-particle systems.
Finally, we present a comprehensive analysis of our DNS results for many-particle systems. Here the collective motion depends on the magnitude of the induced dipole moments, the area fraction of particles, and the strength of the interparticle attraction.

Many computational simulation methods that include HIs have been proposed for systems of microswimmers, including colloidal rollers \cite{shaebani2020computational}.
One example is the smoothed profile (SP) method. This has been applied to study the structure and dynamics in a variety of particle dispersions, such as colloidal systems and self-propelled microswimmers \cite{yamamoto2021smoothed}.
The SP method can describe both near-field and far-field hydrodynamic effects reasonably accurately and is applicable to dense systems \cite{nakayama2005simulation, nakayama2008simulating}.
Here, we use dimensionless simulation units throughout the length given by the grid spacing, the time and the energy by the fluid density and the fluid viscosity, and the electric dipole by the dielectric constant in vacuum (see Methods for more details).

First, as an evaluation of the SP method, we confirm that the speed of the single roller is in good agreement with lubrication theory \cite{goldman1967slow, cichocki1998image}, and that the flow field around the roller agrees with previous simulations that used a rotlet representation corrected by the Blake tensor \cite{blake1974fundamental, balboa2017hydrodynamics, driscoll2017unstable, delmotte2019hydrodynamically} (see Supplementary Information).
The advantage of the SP method is that it can accurately handle HIs between rollers. We first focus on how the cluster states of two and three rollers are affected by these lubrication effects before later studying many-body systems.

\begin{figure*}[htbp]
  \centering
  \includegraphics[clip, width=16.0cm]{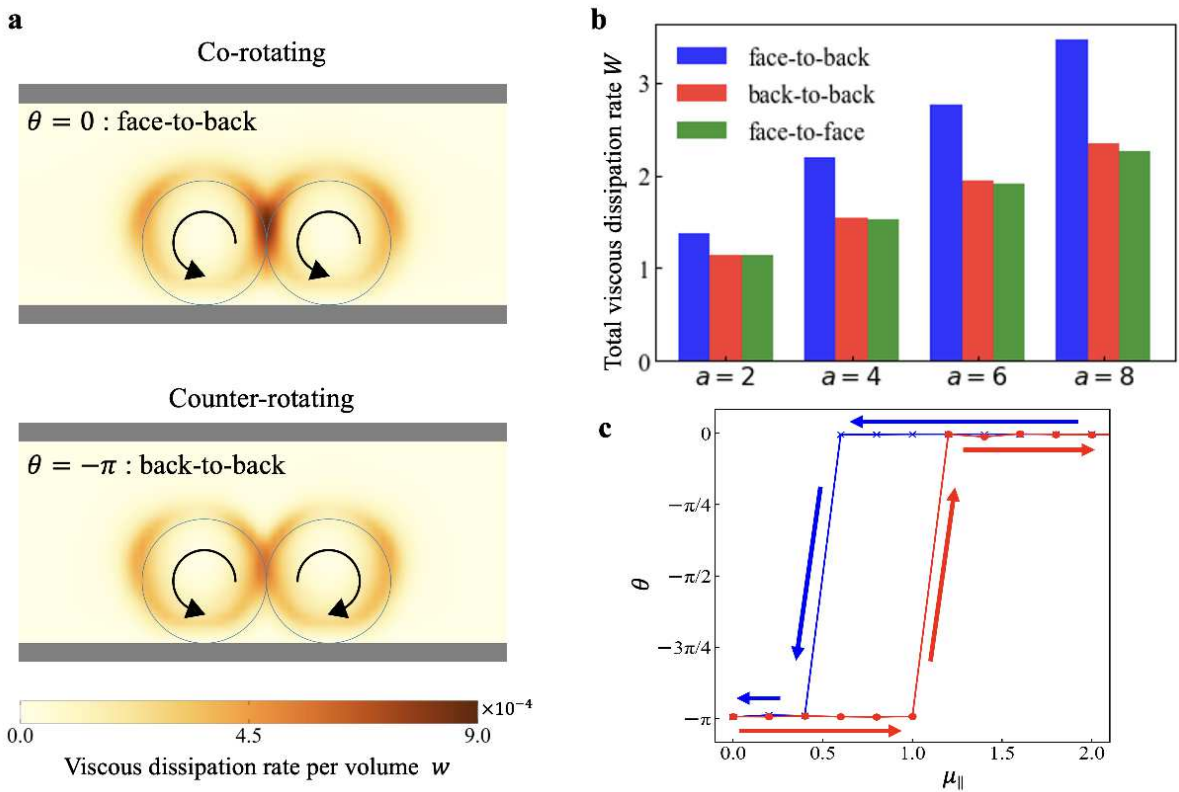}
\caption{
Behavior of a cluster consisting of two rollers.
(a) The viscous dissipation rate per volume, $w = [(\bm{\nabla}\bm{u}) + (\bm{\nabla}\bm{u})^{T}]^2/2$, for the face-to-back ($\theta = 0$) and back-to-back ($\theta = -\pi$) states for radius $a = 8$ particles, and the blue circles represent the boundary line of each particle. The viscous dissipation in the space between the two rollers is higher than it is elsewhere, and the dissipation in the face-to-back state is larger than that in the back-to-back state.
(b) The counter-rotating (face-to-face or back-to-back) states dissipate less heat and might therefore be expected to be more stable than the co-rotating (face-to-back) state, based on the principle of minimum energy dissipation. Here, the face-to-face (back-to-back) state represents the case in which the two rollers are being propelled toward (away from) each other. Results shown for different roller sizes $a = 2$, $4$, $6$, and $8$.
(c)
Stable steady state for different dipole strengths draws a hysteresis curve, where $a = 2$.
Arrows and colors in this curve indicate the direction of the loop.
The HI stabilizes the back-to–back motion $\theta = -\pi$ at $\mu_{\parallel}$ is small,
and the electrostatic interaction stabilizes the face-to-back motion $\theta = 0$ at $\mu_{\parallel}$ is large.
Here, the strength of attraction sets $\epsilon = 10$, and all units are dimensionless, see Methods for more details.
}
  \label{fig: two rollers}
\end{figure*}

The stable cluster states for two-body systems can be classified into a co-rotating (face-to-back) state, in which the rollers are propelled in the same direction, $\theta = 0$, and a counter-rotating (face-to-face $\theta = \pi$ or back-to-back $\theta = -\pi$) state, in which they rotate in opposite directions, where $\theta$ is the relative angle between the axes of Quincke rotation of the rollers (see Supplementary Movie 1).
In order to understand the stabilizing structure of clusters by HI, we consider the viscous dissipation of the system. The viscous dissipation of the fluid around the clusters in the co-rotating and counter-rotating states is shown in \Figref{fig: two rollers} (a). The viscous dissipation rate is defined as $W = \int w(\bm{r}) d\bm{r}$ where the density $w(\bm{r}) = \left[(\bm{\nabla}\bm{u}(\bm{r})) + (\bm{\nabla}\bm{u}(\bm{r}))^{T}\right]^{2} / 2$. According to the principle of minimum energy dissipation \cite{strutt1871some, onsager1931reciprocal1, onsager1931reciprocal2}, a nonequilibrium system prefers to be in a state with lower energy dissipation.
In this system, the viscous dissipation in a counter-rotating state is lower than that in the co-rotating state, see \Figref{fig: two rollers} (b). In addition, from the discussion regarding linear stability \cite{lauga2020fluid} and the forces acting between the rollers, we find that the back-to-back state is more stable than the face-to-back and face-to-face states due to the HI (see Supplementary Information).
On the other hand, the face-to-back state is stabilized by electrostatic interactions at the larger dipole strength. This transition between face-to-back and back-to-back exhibits hysteresis-like behavior surprisingly, see \Figref{fig: two rollers} (c).

\begin{figure}[htbp]
  \centering
  \includegraphics[clip, width=8.0cm]{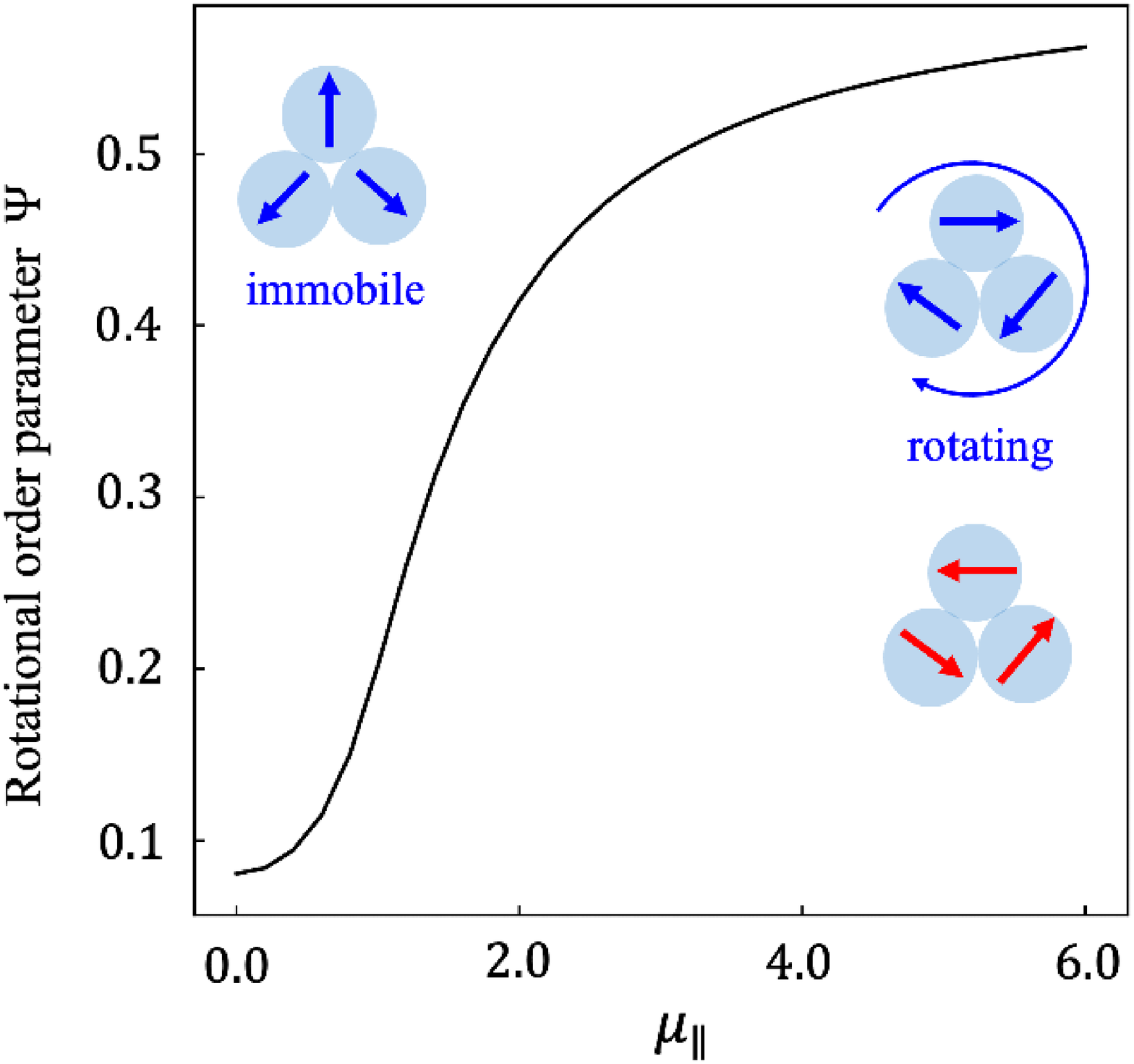}
\caption{
The rotational order parameter $\Psi$ in a three-body system.
When the dipole strength $\mu_{\parallel}$ is small, the HI propels the particles outward from each other, stabilizing the immobile state of the cluster.
On the other hand, when $\mu_{\parallel}$ is large, the dipole interaction stabilizes the rotating cluster state.
}
  \label{fig: three rollers}
\end{figure}

The behavior of the three-roller cluster can be explained by extending the analysis of the two-roller case.
When the dipole interaction is weak, the HI stabilizes the immobile state, but when the dipole interaction is strong, the rotating cluster state is stabilized (see \Figref{fig: three rollers} and Supplementary Movie 2).
This is a result of the dipole interaction stabilizing the rotation axis of each particle either inward or outward relative to the geometric center of the cluster.
To evaluate this quantitatively, we introduce a rotational order parameter for each cluster,
$\Psi=\braket{|\bm{n}_{i} \cdot \widehat{{\bm{r}}'}_{i}|}$, where $\braket{\cdots}$ denotes an average over the particles and time, the vector symbol $\widehat{\cdot}$ represents a unit vector, the unit vector $\bm{n}_{i}$ refers to the axis of Quincke rotation for the $i$-th roller and ${\bm{r}}'_{i} = \bm{r}_{i} - \bm{r}_{G}$, with $\bm{r}_{G}$ being the center of mass of the cluster (see Supplementary Information).
A continuous transition is observed for the three-body cluster, between the immobile disordered clusters and the rotating clusters.

\begin{figure*}[htbp]
  \centering
  \includegraphics[clip, width=16.0cm]{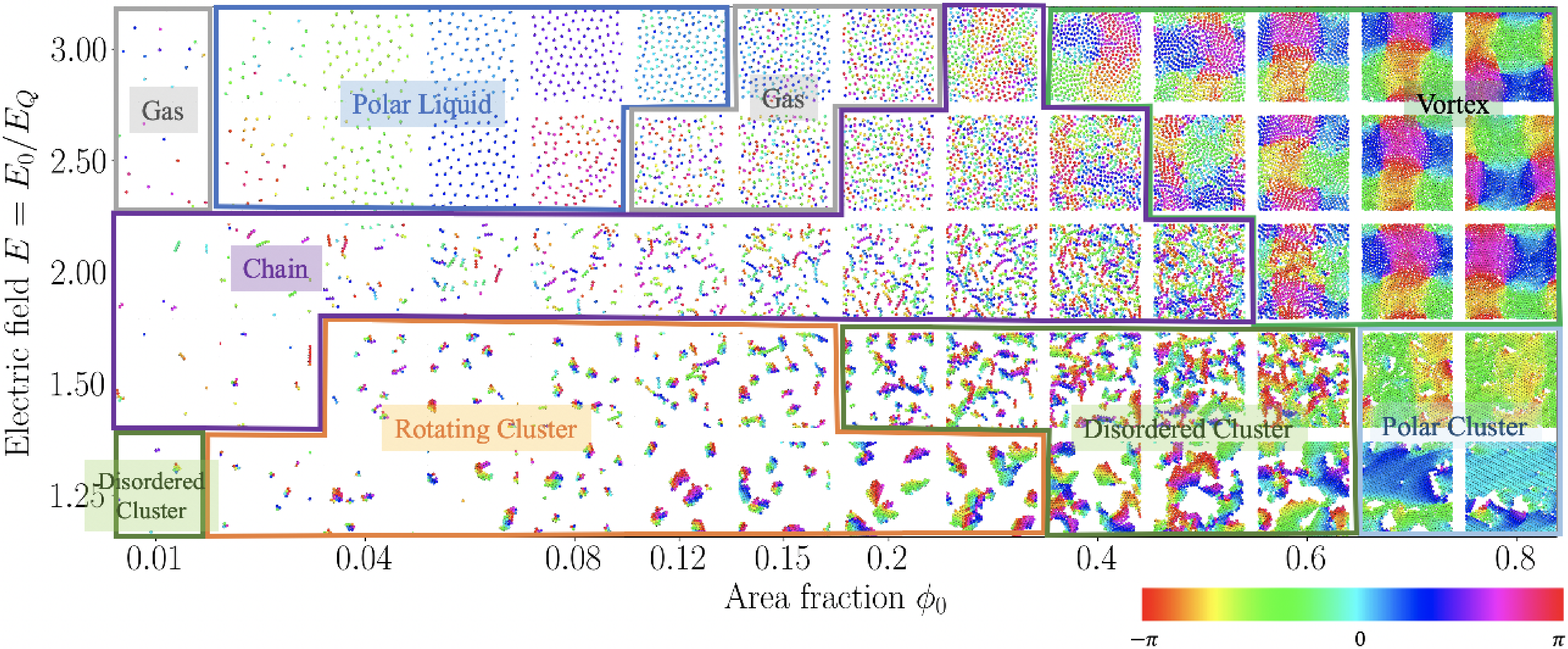}
\caption{
Phase diagram of a many-roller system.
The blue-outlined region represents polar liquid states, the gray-outlined region represents gas states, the green-outlined region represents vortex states, the purple-outlined region represents chain states, the orange-outlined region represents rotating cluster states, the olive-outlined region represents disordered cluster states, and the azure-outlined region represents polar cluster states. See Supplementary Movies 3 and 4 for representative examples of all of these phases.
}
  \label{fig: many rollers}
\end{figure*}

In our simulations for a many-roller systems, we use steady-state solutions for the dipole and angular velocity, which depend on the electric field $E \equiv E_{0}/E_{Q}$\cite{das2013electrohydrodynamic, bricard2013emergence} (see Methods).
The collective structures realized by the many-roller system can be classified by a set of order parameters \cite{imamura2021modeling} as follows: polar liquid states (blue-outlined region), gas states (gray), vortex states (green), chain states (purple), rotating cluster states (orange), disordered cluster states (olive), and polar cluster states (azure) (see \Figref{fig: many rollers}, Supplementary Information and Supplementary Movies 3 and 4).
When the applied electric field $E$ is small, each cluster is a rotating cluster state in the lower-density region, but as the density increases, the clusters tend to become more anisotropic, colliding and coalescing repeatedly to form disordered clusters. At high enough density, all rollers form a single polar cluster.
On the other hand, when the applied electric field $E$ is large, as the area fraction increases, we can find a transition from an ordered polar state to a disordered gas or vortex state.
When the repulsive force exerted by $\mu_{\perp}$ and the attractive force are of the same order, the dipole--dipole interaction given by $\mu_{\parallel}$ dominates, and a chain state is formed.

Our new three-dimensional direct numerical simulation provides important insights into the relationship between the interactions and the collective dynamics.
First, the differences between our observed phase-diagram and previous simulation and experimental results can mainly be traced to the breaking of polar liquid states at higher density\cite{bricard2013emergence, karani2019tuning, mauleon2020competing}.
As the applied electric $E_{0}$ field increases, the repulsion due to $\mu_{\perp}$ exceeds the attraction due to the electro-osmotic flow, causing the rollers to move apart.
Then, as the area fraction $\phi_{0}$ increases, the average distance between the rollers changes, balancing the HIs and dipole interactions and causing a transition from a disordered gas state to an ordered polar liquid state, similar to the results by Bricard {\it et al.} \cite{bricard2013emergence}.
As the area fraction $\phi_{0}$ increases further, the interactions become unbalanced again, and transitions to the disordered gas and vortex states from polar liquid states are observed at $\phi_{0} \geq 0.2$ (see Supplementary Information).
Our results suggest that the disordered dynamics of the active solid state observed by Geyer {\it et al.} \cite{geyer2019freezing} are caused by the polar state disappearing at high density due to the lubrication effects.
However, unlike in the experiments, the solidification from polar liquid states has not been observed in our simulations, which we believe is due to implementation differences, e.g. different boundary conditions, system size and number of particles.
Second, we expect our simulations to reach the chain state whenever the dipole $\mu_{\parallel}$ becomes dominant, in where isotropic forces between the electrostatic forces by $\mu_{perp}$ and electro-osmotic attractive forces are balanced at $E \sim 2.0$. Such a chain state has been observed actually under different conditions with the proceeding researches\cite{bricard2013emergence, mauleon2020competing}, including larger colloid size and different concentration of conducting liquid experimentally\cite{lopez2020important}, and the active Brownian particles with dipole--dipole interactions numerically\cite{liao2020dynamical}.
Finally, they provide new insights into the physical mechanisms responsible for various cluster motions observed in previous experiments \cite{karani2019tuning, mauleon2020competing, liu2021activity}. When the applied electric field $E_{0}$ is small, the attraction due to the electro-osmotic flow is dominant and the cluster motion is reproduced. A rotating cluster created by stabilization through $\mu_{\parallel}$, and clusters with anisotropic shapes (e.g., elliptical), which tend to form at high densities, as well as select translational motion in a polar cluster state. A disordered cluster arises due to the lubrication effect and the shape change caused by frequent coalescence and separation between clusters.

In conclusion, we take advantage of DNSs to include HI and investigate the flocking transition mechanism of Quincke rollers and their dynamics in high-density regions.
The DNS of the fluid flow is performed using the SP method.
Our simulations reveal that the transition between cluster states is caused by the balance between the disordering of the orientation due to the lubrication effects and the ordering of the orientation due to the horizontal component of the electric dipole, and the deformation of the cluster shape.
Additionally, our simulations show an intermediate transition to disordered states from polar liquid states at high densities, due to an imbalance between hydrodynamic and dipole interactions.
Our proposed DNS method can be easily extended to the collective dynamics of colloidal rollers in magnetic fields \cite{kaiser2017flocking, driscoll2017unstable}.
Our findings show that the HIs differ between the near and far fields, which indicates the possibility of new improvements in particle-based models.

\subsection*{Acknowledgements}
This work was supported by Grants-in-Aid for Scientific Research (JSPS KAKENHI) under grant nos. JP 20H00129, 20H05619, and 20K03786.
It was also supported by Professional development Consortium for Computational Materials Scientists (PCoMS).
R.Y. acknowledges helpful discussions with Profs. Hajime Tanaka and Akira Furukawa.

\subsection*{Author contributions}
All authors designed the research and wrote the paper. S.I., K.S. and R.Y. performed the simulations and analysed the data. J.J.M. assisted with numerical methods.

\subsection*{Competing interests}
The authors declare no competing interests.

\subsection*{Additional information}
Supplementary Information is available for this paper at [url will be inserted by publisher].

\clearpage

\subsection*{Methods}
\subsubsection*{Smoothed Profile (SP) method}
The presence of the $i$-th particle is described by the SP function $\phi_{i}$, which takes values of $\phi_{i} = 1$ in the solid domain and $\phi_{i} = 0$ in the fluid domain. These domains are smoothly connected by an interface of finite width $\xi$ at the particle surface \cite{nakayama2005simulation}.
In the SP method, the velocity field $\bm{u}$ is defined at all positions $\bm{x}$ and times $t$ so as to interpolate between the host fluid, moving with velocity $\bm{u}_{f}$, and the solid particles moving with velocity $\bm{u}_{p}$,
\begin{equation}
\bm{u}(\bm{x}, t) = (1 - \phi) \bm{u}_{f}(\bm{x}, t) + \phi \bm{u}_{p}(\bm{x}, t). \label{eq:u}
\end{equation}
In the equation above, the particle field $\phi$ is to be treated as a sum over the contribution of all particles. The two terms in \Braref{eq:u} are  defined as
\begin{equation}
(1-\phi) \bm{u}_{f}(\bm{x}, t) = \left(1-\sum^{N}_{i = 1} \phi_{i}(\bm{x}, t)\right) \bm{u}_{f}(\bm{x}, t)
\end{equation}
and
\begin{equation}
\phi \bm{u}_{p}(\bm{x}, t) = \sum^{N}_{i = 1} \phi_{i}(\bm{x}, t) (\bm{V}_{i} + \bm{\Omega}_{i} \times \bm{r}_{i})
\end{equation}
where $\bm{r}_{i} = \bm{x} - \bm{R}_{i}$ is a radial vector originating at the center of the $i$-th particle, with $\bm{R}_{i}$ that particle's center, $\bm{V}_{i}$ its velocity, and $\bm{\Omega}_{i}$ its angular velocity. The total particle phase-field $\phi(\bm{x}, t)$ is defined as the superposition of the SP functions of all particles.
The total velocity field is governed by a modified Navier--Stokes equation, together with the incompressibility condition:
\begin{align}
	\rho_{f} (\partial_{t} + \bm{u} \cdot \bm{\nabla}) \bm{u} &= - \bm{\nabla}p + \eta_{f} \bm{\nabla}^2 \bm{u} + \rho_{f} \phi \bm{f}_{p}, \\
	\bm{\nabla} \cdot \bm{u} &= 0,
\end{align}
where $\rho_{f}$ and $\eta_{f}$ are the density and viscosity of the fluid, respectively, and $p$ is the pressure. The rigidity of the particle domains is maintained by the force density field $\phi\bm{f}_{p}$, which is computed to ensure momentum conservation between the fluid domain and the rigid-body domain of the particles \cite{nakayama2005simulation}.

The motion of the Quincke rollers is determined by the Newton--Euler equations of motion:
\begin{align}
\frac{d\bm{R}}{dt} &= \bm{V}, \\
M \frac{d\bm{V}}{dt} &= \bm{F}^{H} + \bm{F}^{P} + \bm{F}^{E}, \\
\dot{\bm{Q}} &= \mathrm{skew}(\bm{\Omega}) \cdot \bm{Q}\\
\bm{I} \cdot \frac{d\bm{\Omega}}{dt} &= \bm{N}^{H} + \bm{N}^{P} + \bm{N}^{C},
\end{align}
where $M = 4 \pi a^{3} \rho_{p} /3 $ is the mass of a particle, $\bm{Q}$ the particle's rotation matrix ($\mathrm{skew}(\bm{\Omega})$ the skew-symmetric angular velocity matrix), $\bm{I} = 2 a^{2} M \mathsf{I} /5$ the moment of inertia ($\mathsf{I}$ is the unit tensor), $a$ the particle radius and $\rho_{p}$ the particle density. The hydrodynamic forces and torques due to the fluid-particle interactions are denoted as $\bm{F}^{H}$ and $\bm{N}^{H}$, respectively \cite{nakayama2005simulation}.
The forces and torques due to the inter-particle interactions, $\bm{F}^{P}$ and $\bm{N}^{P}$, are derived from an interaction potential $U = U^{S} + U^{A} + U^{D}$, which is assumed to be pairwise additive\cite{allen2006expressions} (see Supplementary Information). The corresponding potential between a pair of particles $i$ and $j$, with dipole moments $\bm{P}_i$ and $\bm{P}_j$ and separation $\bm{r}=\bm{R}_j - \bm{R}_i$ is 
\begin{align}
U_{ij}^{S}(r) &= 4 \epsilon^{LJ} \left[ \left( \frac{\sigma}{r} \right)^{36} - \left( \frac{\sigma}{r} \right)^{18} \right] + \epsilon^{LJ}, \\
U_{ij}^{A}(r) &= - \epsilon^{A} \exp{(-\kappa r)} / r^{2}, \\
U_{ij}^{D}(\bm{r}, \bm{P}_i, \bm{P}_j) &= \frac{1}{4\pi\epsilon_{0}} \left[ \frac{\bm{P}_{i} \cdot \bm{P}_{j}}{r^3} - 3 \frac{(\bm{P}_{i} \cdot \bm{r})(\bm{P}_{j} \cdot \bm{r})}{r^{5}} \right].
\end{align}
Here the $U_{ij}^{S}$ potential represents the excluded volume effect, and is described by a truncated $18-36$ Lennard--Jones potential, with a cut-off at $r_{c}=2^{1/18} \sigma$, with $\sigma = 2a$ the particle diameter.
The $U_{ij}^{A}$ potential is used to describe the electro-osmotic attraction, where the screening length is given by $\kappa = 1/(3\sigma)$ \cite{mauleon2020competing}.
The strengths of the Lennard-Jones repulsion and electro-osmotic attraction are given by $\epsilon^{LJ}$ and $\epsilon^{A}$, respectively.
The $U_{ij}^{D}$ represents the electrostatic dipole--dipole interactions and includes the interactions of the image dipoles \cite{bricard2013emergence}.
To mimic experimental conditions, we set the applied DC electric field to be perpendicular to the walls, i.e., $\bm{E}_{0} = E_{0} \bm{e}_{z}$ where $\bm{e}_{z}$ is an unit vector of the $Z$ axis, and we set also the angular velocity vector of a single roller due to its Quincke rotation $\bm{\Omega}_{0} = \Omega_{0} \bm{n}$, where the initial orientation is randomly chosen to be parallel to the walls ($XY$-plane), i.e., at the time the electric field is applied, where the magnitude $\Omega_0$ depends on the scaled electric field $E_{0}$ (see \Eqref{equ: rotation speed}).
The electric dipole is defined as $\bm{P} = P_{\parallel} \bm{e}_{p} + P_{\perp} \bm{e}_{z}$, where $P_{\parallel}$ and $P_{\perp}$ are the components parallel and perpendicular to the wall, respectively, and $\bm{e}_{p}$ is a unit vector in the direction of $\bm{E}_{0} \times \bm{\Omega}_{0}$.
These electrostatic interactions are calculated via the Ewald method \cite{aguado2003ewald,allen2017computer,dwelle2019constant}.
Here, $P_{\parallel}$ gives the dipole--dipole interaction on the plane, and $P_{\perp}$ gives an isotropic repulsion \cite{bricard2013emergence}.
$\bm{F}^{E}$ represents the (external) forces exerted by gravity and the wall, where the wall--particle interaction is again represented by a truncated Lennard--Jones potential, similar to the particle-particle steric interactions, that is chosen to be strong enough to exclude the particles from the wall.
Finally, to implement the Quincke roller dynamics, we introduce a constraint torque $\bm{N}^{C}$, derived from a potential $U^{C}$\cite{goldstein2002classical}, to fix the rotation denoted by $\bm{\Omega}_{0}$ to lie in the plane (see Supplementary Information).

\subsubsection*{Parameters and Units}
\paragraph{The electric dipole moments and the rotation speed by Quincke rotation for the steady-state solution:}
Let us consider the correspondence between the theoretical and simulation parameters.
In Quincke rotation, in accordance with the time evolution of an electric dipole\cite{das2013electrohydrodynamic, bricard2013emergence}, the steady-state solution for the dipole moment induced in a dielectric sphere can be expressed as follows:
\begin{align}
P_{\perp} = P_{\perp}^{\epsilon} E - \frac{P_{\perp}^{\sigma}}{E},
P_{\parallel} = P_{\perp}^{\sigma} \sqrt{1-\frac{1}{E^2}},
\label{equ: dipoles}
\end{align}
where $E \equiv E_{0}/E_{Q}$, $P_{\perp}^{\epsilon} \equiv 4 \pi \epsilon_{0} a^3 \chi^{\infty} E_{Q}$ is the static contribution from the dielectric polarization, and the $z$ component $P_{\perp}^{\sigma} \equiv \epsilon_{0}/(\tau \epsilon_{l} \mu_{r} E_{Q})$ and the $xy$ component $P_{\parallel}$ are the dynamic contributions from the Quincke rotation for $E > 1$.
$E_{Q}$ is the threshold value for Quincke rotation, $\tau$ is the Maxwell--Wagner time, $\mu_{r} = (8 \pi \eta_{f} a^{3})^{-1}$ is the mobility of the rotational motion, and $\chi^{\infty} \equiv \frac{\epsilon_{p} - \epsilon_{l}}{\epsilon_{p} + 2\epsilon_{l}}$, with $\epsilon_{p}$ and $\epsilon_{l}$ the dielectric permittivities of the sphere and the liquid medium, respectively, and $\epsilon_{0}$ the dielectric permittivity of vacuum.
Additionally, the angular velocity of a roller in the steady state is described as
\begin{align}
\Omega_{0} = \frac{1}{\tau} \sqrt{E^2 - 1}.
\label{equ: rotation speed}
\end{align}
Note that the accuracy of approximation by this steady-state solutions is guaranteed by the fact that the unit of time $t_{0}$ is sufficiently longer than the Maxwell--Wagner time $\tau$, e.g. $t_{0} \gg \tau$ (see the paragraph of units about a definition of the unit $t_{0}$).

\paragraph{Units:}
Within the SP method, the system is discretized on a rectangular grid with spacing $\Delta$, under periodic boundary conditions in all directions. This allows us to employ a pseudo-spectral method, and significantly reduce the computational cost.
The units of all variables are based on the fundamental simulation parameters $\Delta$, $\eta_{f}$ and $\rho_{f}$, the grid spacing, fluid viscosity, and fluid density respectively.
The unit of length is $l_{0} = \Delta$, time $t_{0} = \rho_{f} \Delta^{2} / \eta_{f}$, and that of energy is $e_{0} = \eta_{f}^{2} \Delta / \rho_{f}$. The electric dipole is measured in units of $p_{0} = \sqrt{4 \pi \epsilon_{0} e_{0} l_{0}^{3}}$.
Here, the viscosity of the liquid is $\eta_{f} \sim 4.3 \times 10^{-3}$ Pa $\cdot$ s, the density ratio between the colloids and the liquid is $\rho_{p}/\rho_{f} = 1.19/0.77 \sim 1.5$\cite{pradillo2019quincke}, the dielectric permittivities $\epsilon_{p} \sim 2.6 \epsilon_{0}$\cite{pannacci2007insulating}, $\epsilon_{l} \sim 2.0 \epsilon_{0}$\cite{schmidt2012conductivity}, and the ratio $\chi^{\infty} \sim 0.08$\cite{bricard2013emergence}.
Based on these physical quantities, we can estimate the parameter units to $l_{0} = \Delta = 1$~$\mu$m, $t_{0} = \frac{\rho_{f} \Delta^2}{\eta_{f}} \sim 0.1$~$\mu$s, $e_{0} = \frac{\eta_{f}^2 \Delta}{\rho_{f}} \sim  10^{-14}$~J, and $p_{0} = \sqrt{4 \pi \epsilon_{0} e_{0} l_{0}^{3}} \sim 10^{-21}$ C$\cdot$m.
Because the Maxwell-Wagner time $\tau \sim 1$ ms\cite{bricard2013emergence}, the accuracy of the steady-state solutions \Eqsref{equ: dipoles} and \Braref{equ: rotation speed} is not sufficiently guaranteed in this unit $t_{0}$.
However, to simplify the problem and to build our fundamental understanding in the collective system, we proceed with the simulation under the assumption that the contribution from the steady-state is dominant.

\paragraph{Simulation parameters:}
We set the simulation parameters from these theoretical values, and we define $\mu_{\parallel}$ and $\mu_{\perp}$ as the dimensionless dipole components corresponding to $P_{\parallel}$ and $P_{\perp}$, respectively, where in what follows all units are dimensionless.
The system size is $(L_{x}, L_{y}, L_{z}) = (128, 128, 32)$ and we introduce a flat wall parallel to the $XY$ plane, with a total depth of $d = 5$, i.e., $H$ is fixed for all simulations. 
The particle Reynolds number is defined as $Re \equiv \rho_{f} \sigma^{2} \Omega_{0} / \eta_{f}$, and $Re \sim 10^{-5}$ in experiments\cite{bricard2013emergence, pradillo2019quincke}, where we assume that $Re \sim 1$ to reduce the computational cost.
Because the excluded volume effect is sufficiently larger than the electro-osmotic attraction, we set $\epsilon \equiv \epsilon^{LJ} = \epsilon^{A}$ for simplicity, and we takes the interfacial width of the particle domain $\xi = 2$ with high reliability\cite{nakayama2005simulation}.
For single, two, and three rollers simulation, we set $Re = 1$.
To investigate the stable structures of cluster states in the simulation for two and three rollers, we assume that the strength of repulsive force $\mu_{\perp} = 0$ and the attractive force $\epsilon = 10.0$ for two rollers and $\epsilon = 30.0$ for three rollers to keep cluster motion, where the radius $a = 2,4,6,$ and $8$ for the two rollers and $a = 2$ for the three rollers.
Additionally, for many-rollers simulation, we set $a = 2$, the strength of the attractive force $\epsilon=10.0$, and the area fraction is varied in the range $\phi_{0}=0.01-0.8$, based on the steady-state solution \Eqsref{equ: dipoles} and \Braref{equ: rotation speed}, each input parameter is determined from the electric field $E$ as follows: the angular velocity is $\Omega_{0} \sim 0.025\times\sqrt{E^{2}-1}$, the dipole moment is $\mu_{\parallel} \sim 2.6 \times \sqrt{1 - \frac{1}{E^2}}$, and the ratio between the dipole components is $|\mu_{\perp}|/|\mu_{\parallel}| \sim E$.

\clearpage
\begin{widetext}
\setcounter{page}{1}

\begin{center}
Supplemental Information for the Manuscript \\
``Collective Motions of Quincke Rollers with Fully Resolved Hydrodynamics''\\
\ \\
Shun Imamura, Kohei Sawaki, John J. Molina, Matthew S. Turner, and Ryoichi Yamamoto\\
Department of Chemical Engineering, Kyoto University, Kyoto 615-8510, Japan
\end{center}


\section{Simulation Details} \label{appendix: simulation details}
\subsection{Derivation of forces and torques from the potential}
The potential $U$ depends not only on the relative position of the particles $\bm{r}$ but also on their relative orientations in space, as given by the rotation axis for Quincke rotation and the dipole moments of the particles.
The interparticle forces and torques can be obtained from the potential as follows \cite{allen2006expressions}:
\begin{align}
\bm{F}^{P} &= - \frac{\partial U}{\partial \bm{r}},\\
\bm{N}^{P} &= - \sum_{m} \widehat{\bm{a}}_{m} \times \frac{\partial U}{\partial \widehat{\bm{a}}_{m}},
\end{align}
where the $\widehat{\bm{a}}_i$ are the (body-fixed) ortho-normal principal axis vectors ($m=x', y', z'$).

\subsection{A torque to constrain the Quincke rotation}
\begin{figure}[htbp]
  \centering
  \includegraphics[clip, width=10.0cm]{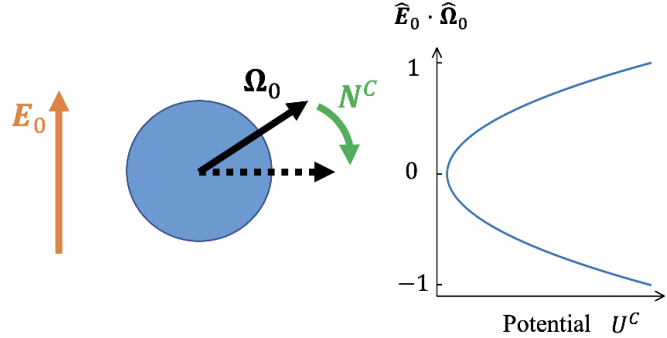}
\caption{
Schematic diagram of the torque used to constrain the axis of rotation.
The angle between the applied electric field $\bm{E}_{0}$ and the angular velocity by the Quincke rotation $\bm{\Omega}_{0}$ is constrained to be perpendicular by a harmonic potential.
}
  \label{fig: constraint}
\end{figure}
We introduce an external torque inspired by a Quincke roller.
It has been analytically proven that the angular velocity vector generated by Quincke rotation is perpendicular to the external electric field $\bm{E}_{0}$\cite{pannacci2007insulating}.
In our model, to simplify the treatment, we impose a constant angular velocity vector $\bm{\Omega}_{0}$ on a rigid sphere and a constraint torque $\bm{N}^{C}$ defined by a harmonic potential is introduced to maintain the rotation axis of the sphere to a plane, see \Figref{fig: constraint}.
Specifically, the constraint potential and torque are defined from the directions of $\bm{E}_{0}$ and $\bm{\Omega}_{0}$ as follows:
\begin{align}
U^{C}
&=
\frac{K}{2} \left( \widehat{\bm{E}}_{0} \cdot \widehat{\bm{\Omega}}_{0} \right)^{2}
,
\\
\bm{N}^{C} &=
K \left( \widehat{\bm{E}}_{0} \cdot \widehat{\bm{\Omega}}_{0} \right)
 \left( \widehat{\bm{E}}_{0} \times \widehat{\bm{\Omega}}_{0} \right).
\end{align}
The vector symbol $\widehat{\cdot}$ represents a unit vector.
$K > 0$ is the amplitude of the constraint torque, whose value must be sufficiently large relative to the hydrodynamic torque $\bm{N}^{H}$
because the axis of rotation should be fixed.
The constraint torque is obtained from the potential as $\displaystyle{ \widehat{\bm{\psi}} \cdot \bm{N}^{C} = - \frac{\partial U^{C}}{\partial \psi} = - \frac{\partial U^{C}}{\partial \widehat{\bm{\Omega}}_{c}} \cdot  \frac{\partial \widehat{\bm{\Omega}}_{0}}{\partial \psi}}$ for angle $\psi$ about any axis represented by a unit vector $\widehat{\bm{\psi}}$, where we use the rotation formula $\displaystyle{\frac{\partial \widehat{\bm{\Omega}}_{0}}{\partial \psi} = \widehat{\bm{\Omega}}_{0} \times \widehat{\bm{\psi}}}$ \cite{goldstein2002classical}.

\subsection{Simulation procedure}
The simulation procedure is as follows:
\begin{enumerate}
\item We obtain $\phi^{n}$ and $\bm{u}^{n}$ from the configuration $\{ \bm{R}_{i}^{n} \}$ and the velocities $\{ \bm{V}_{i}^{n} \}$ and $\{ \bm{\Omega}_{i}^{n} \}$, where the superscript $n$ denotes the time step, and then we calculate $\bm{u}^{*}$ using $\bm{u}^{n}$.
\item We update the configuration $\{ \bm{R}_{i}^{n}, \bm{Q}_i^n\}$ to $\{ \bm{R}_{i}^{n+1}, \bm{Q}_i^{n+1}\}$ using the velocities $\{ \bm{V}_{i}^{n}, \bm{\Omega}_i^n \}$ and obtain the updated $\phi^{n+1}$.
\item We calculate the hydrodynamic force $\bm{F}_{i}^{H}$ and torque $\bm{N}_{i}^{H}$ from the updated profile function $\phi^{n+1}$ and then further calculate the forces $\bm{F}_{i}^{P}, \bm{F}_{i}^{E}$ and torques $\bm{N}_{i}^{P}, \bm{N}_{i}^{C}$ by inter-particle interactions and external field.
\item From the forces and torques, we obtain the updated velocities $\{ \bm{V}_{i}^{n+1} \}$ and $\{ \bm{\Omega}_{i}^{n + 1} \}$; then, we return to step 1.
\end{enumerate}
The equations of motion are solved using the second-order Adams--Bashforth method.
The time step is taken to be $\Delta t = \rho_{f} / (\eta_{f} k_{max}^2)$,
where $k_{max}=2\pi/\Delta$ is the maximum wavenumber.

\subsection{Numerical validations} \label{appendix: numerical validations}
\begin{figure}[htbp]
  \centering
  \includegraphics[clip, width=16.0cm]{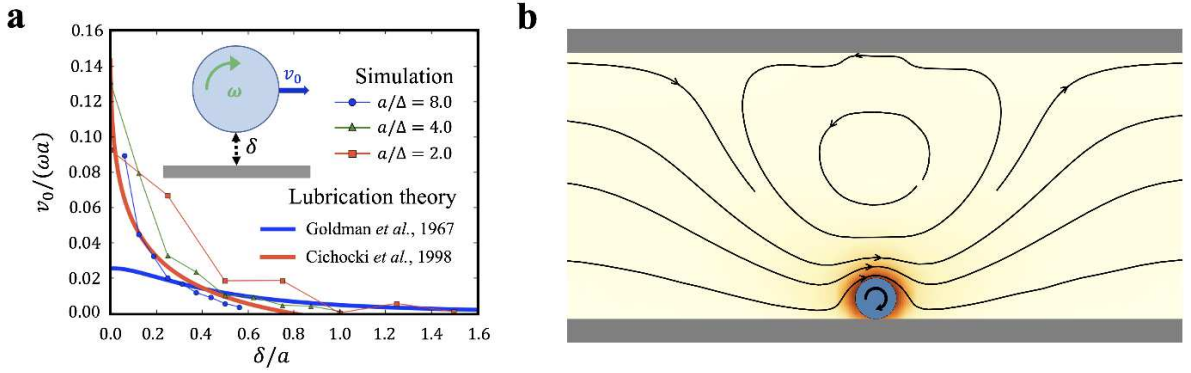}
\caption{
(a) Normalized roller velocity $v_{0}/(a \Omega_{0})$ as a function of the distance between the surface of the particle and the wall, $\delta$, for comparison with lubrication theory (far field: Goldman {\it et al.}, 1967; near field: Cichocki {\it et al.}, 1998).
(b) Flow field around a single roller.
}
  \label{fig: validation}
\end{figure}
We present a numerical validation for the propulsion of a colloidal roller.
The propulsion speed $v_{0}/(\Omega_{0} a)$ as a function of the distance between the surface of the particle and the wall is derived from lubrication theory\cite{goldman1967slow, cichocki1998image} as follows:
\begin{align}
	f^{tr} &= 
	\begin{cases}
		\frac{1}{8} \left( \frac{a}{h} \right)^{4}
		\left( 1 - \frac{3}{8} \frac{a}{h} \right)
		& \left( \frac{a}{h} \ll 1 \right)
	\\
		\frac{2}{15} \ln{ \left( \frac{\delta}{h} \right)}
		+ 0.25726
		+ \frac{86}{375} \left( \frac{\delta}{a} \right)
		\ln{\left( \frac{\delta}{h} \right)}
		& \left( \frac{\delta}{a} \ll 1 \right)
	\end{cases}
	,
	\\
	f^{tt} &=
	\begin{cases}
		- \left[ 1 - \frac{9}{16} \left( \frac{a}{h} \right)
		 + \frac{1}{8} \left( \frac{a}{h} \right)^{3}
		 - \frac{45}{256} \left( \frac{a}{h} \right)^{4}
		 - \frac{1}{16} \left( \frac{a}{h} \right)^{5} \right]^{-1}
		 & \left( \frac{a}{h} \ll 1 \right)
		 \\
		 -\frac{8}{15}  \ln{ \left( \frac{\delta}{h} \right)}
		 + 0.95429
		 - \frac{64}{375} \left( \frac{\delta}{a} \right)
		\ln{\left( \frac{\delta}{h} \right)}
		 & \left( \frac{\delta}{a} \ll 1 \right)
	\end{cases}
	,
\end{align}
where $h = a + \delta$.
We define $F^{t}$ as the drag force acting on the particle for translational motion and $F^{r}$ as the translational hydrodynamic force for a rotating particle; these forces are defined as $F^{tt} = 6 \pi \mu a f^{tt}(\delta) v_{0}$ and $F^{tr} = 6 \pi \mu a^{2} f^{tr}(\delta) \Omega_{0}$.
Since the forces acting on the particle satisfy the force-free condition, $F^{tt} + F^{tr} = 0$, we obtain the relation $v_{0}(\delta)/(\Omega_{0} a) = - f^{tr}(\delta) / f^{tt}(\delta)$.
Our simulations using the SP method can reproduce the results of lubrication theory, and the accuracy increases as the particle radius increases; see \Figref{fig: validation}(a).
This relation has been confirmed in experiments on Quincke rollers by Pradillo {\it et al.}\cite{pradillo2019quincke} and is a sufficiently reliable result.

The flow field around a single roller appears as shown in \Figref{fig: validation} (b).
This flow field reproduces the results of the multiblob method with a rotlet corrected by the Blake tensor\cite{blake1974fundamental, driscoll2017unstable, delmotte2019hydrodynamically}.
Here, we perform computations for a system that accounts for the effects between the upper and lower wall surfaces, which is slightly different from the system treated in the multiblob method.
In the multiblob method, the mobility matrix is computed via the superposition of two-body interactions based on the Stokesian dynamics method\cite{balboa2017hydrodynamics}, whereas in our SP method, the hydrodynamic equations are treated precisely by direct numerical calculation.

\section{Stabilization by the lubrication effect for a cluster of two rollers}

\subsection{Linear stability analysis for reorientation}
\begin{figure}[htbp]
  \centering
  \includegraphics[clip, width=16.0cm]{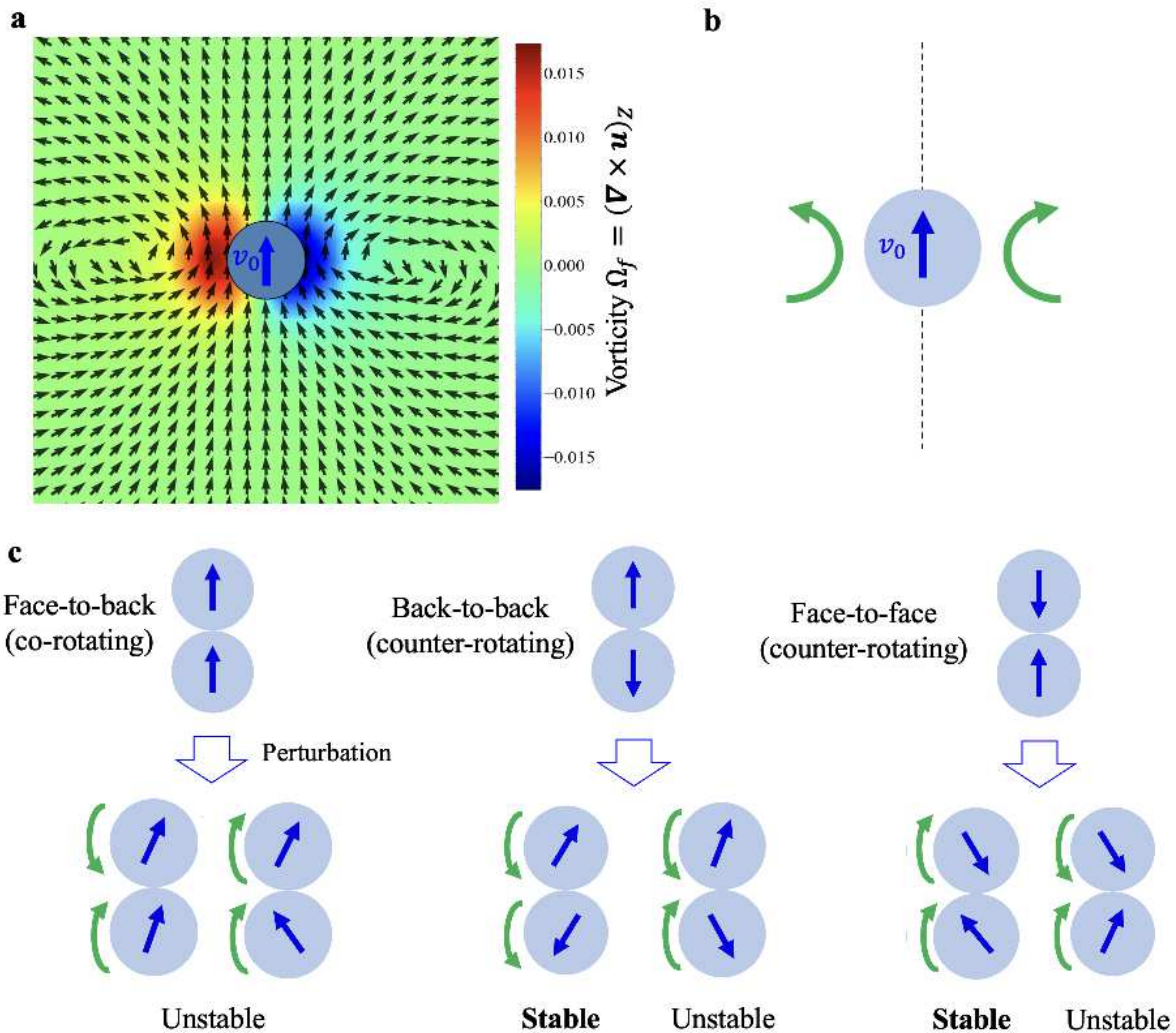}
\caption{
(a) Streamlines and vorticity around a single roller.
(b) The torque around a single roller caused by the lubrication effect, where the vorticity is zero on the dotted line.
(c) The reorientation of the lubrication effect for two rollers clustered by isotropic attraction, where the green arrows show the directions of rotation by hydrodynamic interaction when perturbed.
The direction of perturbation is the same (left side) or opposite (right side) from each roller.
The co-rotating state is unstable, while the counter-rotating (face-to-face and back-to-back) states have configurations that are stable against perturbations.
}
  \label{fig: reorientation}
\end{figure}
In this section, we discuss how a cluster of two rollers, formed by isotropic attraction, is stabilized by the lubrication effect of the HI.
According to Fax{\'e}n's law, due to the HI between the rollers, the rollers are subjected to a torque proportional to the vorticity $\Omega_{f} = \frac{1}{2} \left( \bm{\nabla} \times \bm{u} \right)_{Z}$ caused by the fluid flow\cite{lauga2020fluid}.
The streamlines and the vorticity in the flow field on the plane across the center of gravity of the roller are shown in \Figref{fig: reorientation} (a).
The vorticity is large when the relative distance from the center of gravity of the single roller is approximately equal to the roller diameter, indicating that the influence of the lubrication effect is large.
The torque exerted by the flow field around the single roller based on the vorticity is shown in the schematic diagram in \Figref{fig: reorientation} (b).
As shown in \Figref{fig: reorientation} (c), the fixed point lies on the line of zero vorticity, and there are three possible reorientations: face-to-back (co-rotating), face-to-face (counter-rotating), and back-to-back (counter-rotating) states.
When evaluating the stability of these states against any perturbations, it can be seen that the counter-rotating states are more stable than the co-rotating state.
This stability is consistent with the discussion by the minimum principle of viscous dissipation rate in main text.
However, we cannot know which of the face-to-face and back-to-back states is more stable from this analysis only.

\subsection{Forces acting on two rollers}
\begin{figure}[htbp]
  \centering
  \includegraphics[clip, width=16.0cm]{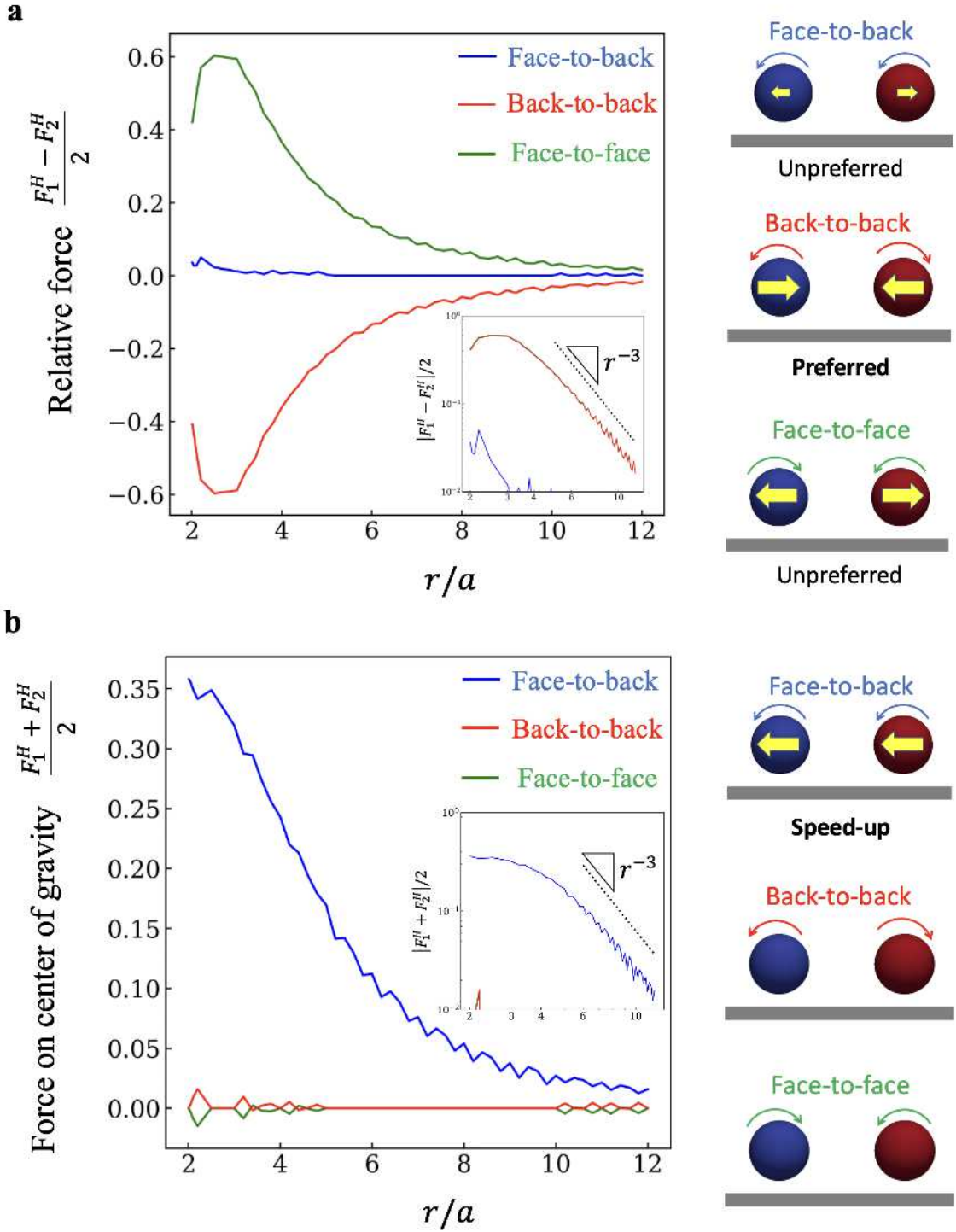}
\caption{
Forces due to the lubrication effect between two rollers.
Blue, red and green lines show the forces acting on the face-to-back, back-to-back and face-to-face states, respectively.
The insets show the log-log plot, where the dotted lines have the slope $r^{-3}$, and the yellow arrows show the directions of the hydrodynamic forces in the schematic illustrations.
(a) The relative force due to the HI, where positive and negative values represent repulsive and attractive forces, respectively.
We can see that the back-to-back state which has a attractive forces differ from others is preferred, the other states have repulsive forces, and the relative force acting on the face-to-back (co-rotating) state is smaller than the forces acting on the other states.
(b) The force acting on the center of gravity due to the HI.
The force acts on same direction in the face-to-back state, accelerating the movement of the two rollers.
}
  \label{fig: two_force}
\end{figure}
To determine the stability and features of the states for two rollers under the lubrication effect, we consider the hydrodynamic forces acting on a cluster formed by two rollers.
The forces $F^{H}_{1}$ and $F^{H}_{2}$ are defined as the forces acting on rollers 1 (blue) and 2 (red), respectively, and the hydrodynamic forces are divided into two components, namely, the relative force $(F^{H}_{1} - F^{H}_{2})/2$ and the force acting on the center of gravity $(F^{H}_{1} + F^{H}_{2})/2$; see \Figref{fig: two_force}.
The relative force acts as an attractive force in the back-to-back state, and this state is preferred under the lubrication effect. In contrast, the relative hydrodynamic force acts as a repulsive force in the face-to-back or face-to-face states, and this states is not preferred under the lubrication effect, as shown in \Figref{fig: two_force}(a).
Only in the face-to-back (co-rotating) state the force act in the direction of propulsion, causing the two rollers to speed up; see \Figref{fig: two_force}(b).
This acceleration phenomenon has been observed experimentally, and it is known that the acceleration effect increases with higher density and higher rotation speed\cite{lu2018pair}.
Thus, the lubrication effect also improves the mobility.

\section{Analysis of many-roller system}
\subsection{Order parameters}
\begin{figure}[htbp]
  \centering
  \includegraphics[clip, width=16.0cm]{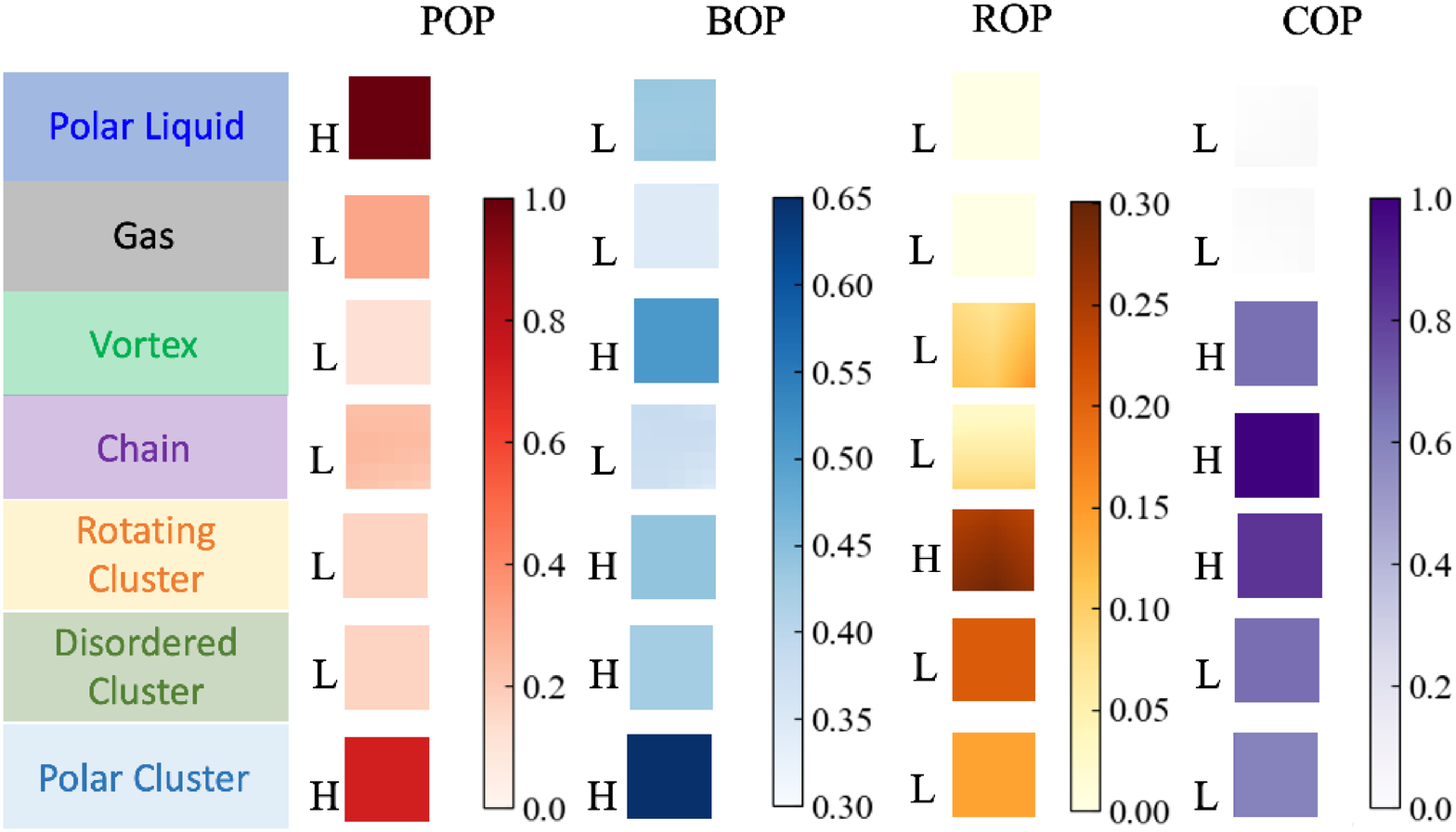}
\caption{
Fingerprint of each state created by the order parameters. The polar order parameter (POP), the bond-orientational order parameter (BOP), the rotational order parameter (ROP) and the chain order parameter (COP) are used to classify the states.
The letters `H' and `L' denote high and low values of the order parameters, respectively, and the threshold values are defined as $0.6$ (POP), $0.40$ (BOP), $0.25$ (ROP) and $0.6$ (COP).
}
  \label{fig: opmaps}
\end{figure}
To characterize each state, we define a polar order parameter (POP), a bond-orientational order parameter (BOP), a rotational order parameter (ROP) and a chain order parameter (COP)\cite{imamura2021modeling}, and we show the corresponding fingerprints of each state (see \Figref{fig: opmaps}).
Note that the POP is defined as the average of the propulsion directions of all particles, $\left| \braket{\bm{n}_{i}} \right|$, where $\bm{n}_{i}$ shows the unit vector of rotation axis by Quincke rotation of the $i$-th roller and $\braket{\cdots}$ denotes an average over the particles and time, while the BOP and COP reflect whether the particles are arranged in a hexagonal configuration or a straight line.
The BOP is defined as $\braket{|\psi_{i}|}$,
\begin{align}
\psi_{i} = \sum_{j = 1}^{Z_{i}} \exp{(i 6 \theta_{ij})}/Z_{i},
\end{align}
where $Z_{i}$ is the coordination number of the $i$-th particle obtained from a Voronoi construction for the particle configuration and $\theta_{ij}$ is the angle between a reference axis and the bond between the $i$-th particle and its $j$-th neighbor.
$\psi = 1$ indicates perfect hexagonal ordering, whereas a completely disordered structure corresponds to $\psi = 0$.
The ROP is defined as $\braket{|\braket{\hat{\bm{\Omega}}_{0,i} \cdot \hat{{\bm{r}}'}_{i}}_{c}|}$, where $\braket{\cdots}_{c}$ denotes taking the average for the particles that make up each cluster in each snapshot and the operation $\braket{\cdots}$ denotes taking the average over the values for each cluster.
${\bm{r}}'_{i} = \bm{r}_{i} - \bm{r}_{G}$, with $\bm{r}_{G}$ being the center of gravity for each cluster, where a cluster is defined as a set of rollers that are in contact with each other.
The COP is defined as $\braket{c_{i}}$,
\begin{align}
c_{i} = \frac{1}{{}_{\mathcal{N}_{i}} C_{2}}
\sum_{(j, k) \in \mathcal{S}_{i}, j \neq k}
\left[
\frac{2}{3}\left( \frac{1}{2} - \frac{\bm{r}_{ij} \cdot \bm{r}_{ik}}{r_{ij}r_{ik}}\right)
\right],
\end{align}
where $M$ represents the number of particles that are in contact with two or more other particles, $\mathcal{N}_{i}$ is the number of particles in contact with the $i$-th particle, and
$\mathcal{S}_{i}$ denotes the region where other particles are in contact with the $i$-th particle, i.e., a circle with a radius on the order of the particle diameter centered on the $i$-th particle\cite{imamura2021modeling}.

\subsection{Two-body correlation function for the angular velocity vector}
\begin{figure}[htbp]
  \centering
  \includegraphics[clip, width=16.0cm]{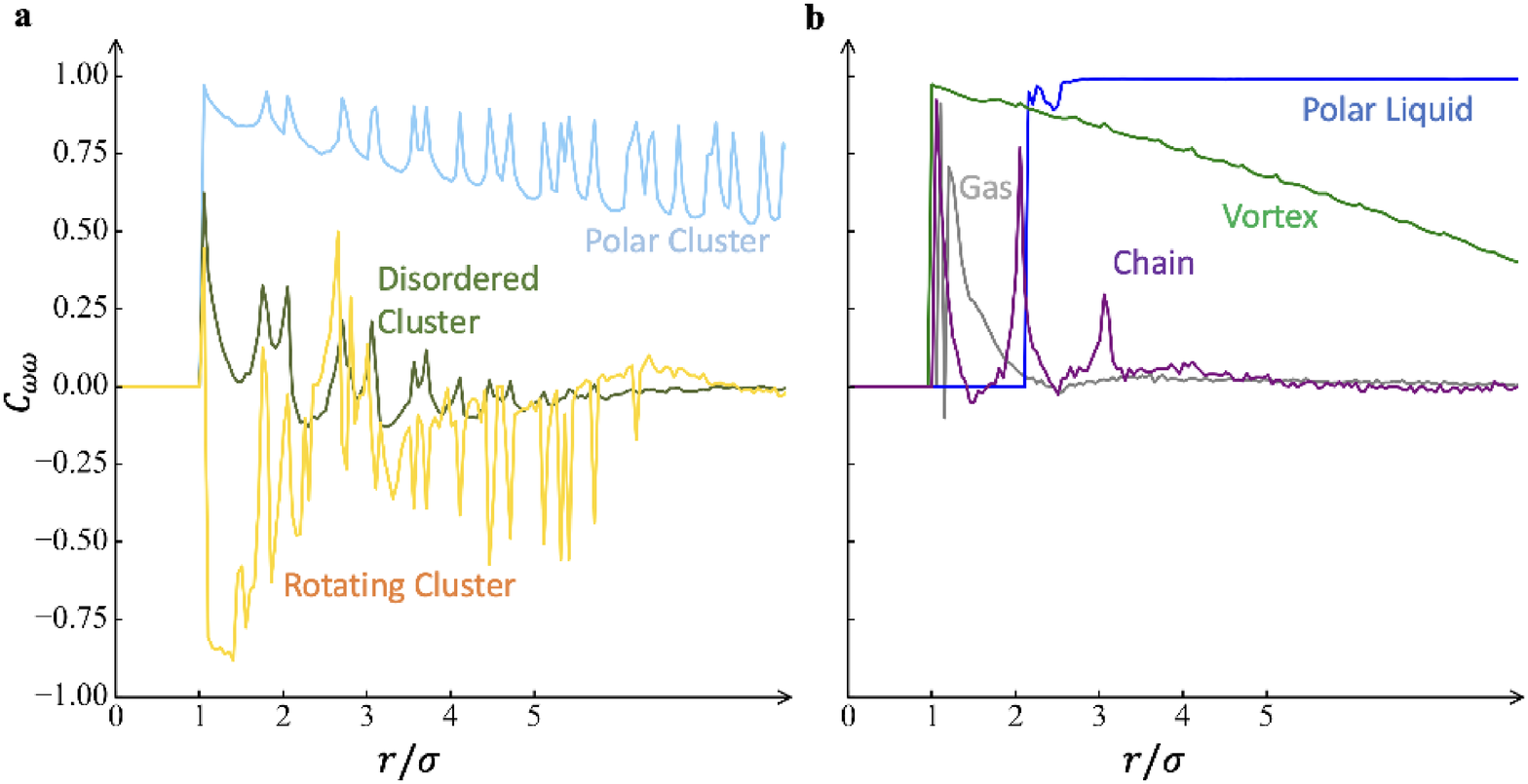}
\caption{
The two-body angular velocity correlations $C_{\omega \omega}(r)$ for each state;
(a) polar, disordered, and rotating cluster states.
(b) gas, polar liquid, vortex, and chain states.
}
  \label{fig: cww}
\end{figure}
Furthermore, to provide an understanding of the characteristic features of each behavior, we introduce the two-body angular velocity correlations as follows:
\begin{align}
    C_{\omega \omega}(r) = \frac{\Braket{\bm{n}_{i} \cdot \bm{n}_{j} \delta(r - r_{ij})}}{2 \pi r \Delta r \phi_{0}(r)},
\end{align}
where $\phi_{0}(r)$ is the local area fraction of the rollers and $2 \pi r \Delta r \phi_{0}(r)$ is the number of rollers in a ring-shaped parcel of width $\Delta r$ and radius $r$.
This function takes values of $C_{\omega \omega} \to 1$ in the parallel state and $C_{\omega \omega} \to -1$ in the antiparallel state.
\Figref{fig: cww}(a) shows functions related to cluster motion, which are characterized by larger fluctuations than those observed in \Figref{fig: cww}(b) for motion under high applied electric fields.
In particular, the rotating cluster function is characterized by a negative peak around the particle diameter and oscillating between positive and negative values, as in \Figref{fig: cww}(a).
The function for chain state forms peaks at intervals roughly corresponding to the diameter scale $2a$, as in \Figref{fig: cww}(b), while the function in the vortex state shows a maximum directional order at the diameter and then decays as a longer distance between rollers.
In the vortex state, the directional order is maximal at the diameter and then decays with distance.
It can be seen that the disorder of the disordered cluster and gas states and the order of the polar clusters and liquids are manifested as the correlations in the long-distance.

\subsection{The breaking of the polar state for denser region}
\begin{figure}[htbp]
  \centering
  \includegraphics[clip, width=12.0cm]{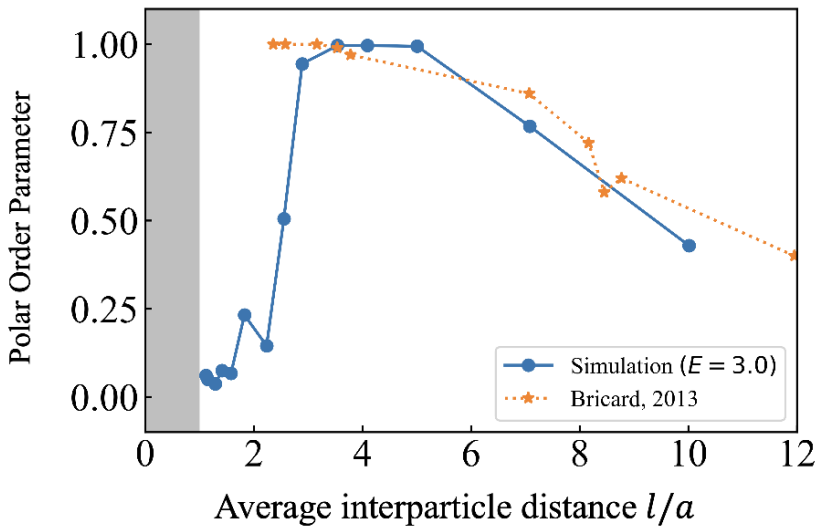}
\caption{
The POP depends on the density of the rollers at $E=3.0$.
The average distance occupied each roller is defined as $l/a \equiv 1/\sqrt{\phi_{0}}$, and the shaded area represents the region inside the roller, $l/a \leq 1$, where $\sigma$ is the diameter of the particle.
The dotted line represents the experimental findings of Bricard {\it et al.}, 2013 at $E=1.39$, and it is seen that our simulation results reproduce the experiments in the lower-density region.
On the other hand, in the higher-density region, the ordering of the flocking is rapidly broken due to the lubrication effect of the HI as the average radius of the region occupied by a single particle becomes smaller than $l/a \sim 2$.
}
  \label{fig: breaking}
\end{figure}
Here, we observe that the POP depends on the density of the rollers, and our results reproduce the experiments\cite{bricard2013emergence} in the lower-density region.
We consider the characteristic distance between particles at which the polar state breaks down at the higher-density region, i.e., the characteristic distance at which the dipole--dipole interaction and the HI lubrication effect switch in dominance.
The average area occupied by each roller is $\bar{S} \equiv L^2 / N = \pi a^2 / \phi_{0}$, and we assume that the shape of this region is circular and define the characteristic distance as $l \equiv \sqrt{\bar{S}/\pi} = a/\sqrt{\phi_{0}}$.
As $l/a$ decays with increasing density, the POP rapidly becomes disordered below $l/a \sim 2$, as shown in \Figref{fig: breaking}.
This characteristic distance, which is equivalent to the diameter scale $2a$, corresponds to the range where the lubrication effect has a large influence on the interparticle interaction and is considered to represent the distance at which the dipole--dipole interaction and lubrication effects switch in dominance in a many-roller system.
As the result, the Lubrication effect of HI is thought to promote further disordering at the higher-density region for a many-roller system.

\section{Descriptions of Supplementary Animations}
We provide animations capturing the time evolution of each state, as obtained from simulations.
\begin{description}
\item[Supplementary Movie 1]
The motions of a two-roller can be classified into two types: co-rotating ($\theta = 0$) and counter-rotating ($\theta = -\pi, \pi$) states.
In the co-rotating state, the cluster moves in one direction at a faster speed than that of a single roller.
On the other hand, in a counter-rotating state, the cluster is immobile because each roller moves in the opposite direction of the other.
In these animations, the color bar represents the normalized amplitude of the velocity field, and the arrows represent the direction of the velocity field at each point.

\item[Supplementary Movie 2]
When the electric dipole moment strength is small, the axes of rotation are disordered by the lubrication effect. As a result, the cluster is immobile.
On the other hand, when the dipole strength is large, the cluster starts to rotate due to the stabilization of the rotational axes by the dipole--dipole interaction.
The rollers are colored in accordance with the direction of the rotational axis, and the arrows also show the direction.

\item[Supplementary Movie 3]
This diagram was obtained from simulations run with parameters corresponding to the theory in the steady state.
The colors of the frames represent the types of states; blue indicates polar liquid states, gray indicates gas states, green indicates vortex states, purple indicates chain states, orange indicates rotating cluster states, olive indicates disordered cluster states, and azure indicates polar cluster states.

\item[Supplementary Movie 4]

A gas state, in which each roller is moving randomly.
The gas states are thought to be caused by an imbalance between HI and dipole interactions between the rollers. The long-range HI is dominant at low densities, while the short-range dipole interaction is dominant at high densities.
Parameters: $E = 3.0$ and $\phi_{0} = 0.01, 0.2$.

A polar liquid state, in which many rollers move in one direction and their distances from each other remain almost constant during the motion.
Isotropic repulsion causes the rollers to separate from each other because of the dominant interaction of $\mu_{\perp}$.
The rollers are not scattered, but are regularly arranged due to the periodic boundary condition.
At the right density, the distance between the rollers is just enough to balance the interaction between the horizontal components of HI and dipole, which is thought to produce directional order.
Parameters: $E = 3.0$ and $\phi_{0} = 0.08$.

A vortex state, in which all the rollers generate enough vortex to cover the entire system.
The average distance between the rollers is almost of the order of the diameter, and not only the electric dipole interaction but also the lubrication effect is strongly acting.
As a result, the torque due to the lubrication effect gradually bends the propulsive direction of the neighboring particles, which is thought to create a vortex.
Parameters: $E = 3.0$ and $\phi_{0} = 0.8$.

A chain state, in which the front and back of the rollers are attached to each other.
Because the isotropic attraction due to the electrohydrodynamic flow is balanced by the isotropic repulsion by $\mu_{\perp}$, $\mu_{\parallel}$ is dominant, creating a chain.
Since the parallel arrangement of rollers like a chain is one of the fixed points in the lubrication effect, the chain state is maintained as long as there is no perturbation due to motion more intense.
Parameters: $E = 2.0$ and $\phi_{0} = 0.08$.

A rotating cluster state, in which each cluster rotates clockwise or counterclockwise.
The rotational motion of the cluster is caused by the stabilization of each roller's rotational axis by the dipole--dipole interaction.
Parameters: $E = 1.25$ and $\phi_{0} = 0.08$.

A disorder cluster state, in which there is no order regarding the direction of propulsion and rotation for the entire cluster.
There are two possible mechanisms for the generation of these disordered clusters.
First, when the dipole is small and the HI is dominant, the cluster becomes disordered due to the lubrication effect.
On the other hand, as higher density, the proportion of clusters that have anistropic shape when the dipole is large increases, and the rotational order within the clusters due to the dipole is decreased.
As a result, each cluster makes random translational and rotational motions, which repeatedly coalesce and separate, creating a disordered state as a whole.
Parameters: $E = 1.25$ and $\phi_{0} = 0.5$.

A polar cluster state, in which all the rollers are moving in one direction.
As higher density sufficiently, all the rollers come together to form a large cluster that spans the boundaries of the system.
Because of the anisotropic shape of this cluster, the rotational order is small and the motion is almost translational.
Parameters: $E = 1.25$ and $\phi_{0} = 0.8$.

\end{description}

\end{widetext}

\bibliographystyle{naturemag}


\end{document}